\documentclass[aps,prb,twocolumn,amsmath,amssymb]{revtex4}
\usepackage{graphicx}
\usepackage{dcolumn}
\usepackage{bm}
\usepackage{here}

\begin{document}

\title{Zero-Temperature Properties of the Quantum Dimer Model on the Triangular Lattice}

\author{Arnaud Ralko,$^1$ Michel Ferrero,$^2$ Federico Becca,$^2$ 
Dmitri Ivanov,$^1$ and Fr\'{e}d\'{e}ric Mila$^1$} 

\affiliation{$^1$Institut de Th\'{e}orie des Ph\'{e}nom\`{e}nes Physiques, 
Ecole Polytechnique F\'{e}d\'{e}rale de Lausanne (EPFL), 
CH-1015 Lausanne, Switzerland\\
$^2$INFM-Democritos, National Simulation Centre and International School 
for Advanced Studies (SISSA), Via Beirut 2-4, I-34014 Trieste, Italy}

\date{February 11, 2005}
             %%%%%%%%%%%%%%%%%%%%%%%%%%%%%%%%%%%%%%%%%%%%%%%%%%%%%%%
             % Do not forget to put the correct submission date!!! %
             %%%%%%%%%%%%%%%%%%%%%%%%%%%%%%%%%%%%%%%%%%%%%%%%%%%%%%%

\begin{abstract}
Using exact diagonalizations and Green's function Monte Carlo simulations,
we have studied the zero-temperature properties of the quantum dimer model
on the triangular lattice on clusters with up to 588 sites. A detailed comparison
of the properties in different topological sectors as a function of the cluster size and for
different cluster shapes has allowed us to identify different phases,
to show explicitly the presence of topological degeneracy in a phase close to the 
Rokhsar-Kivelson point, and to understand finite-size effects inside this phase. 
The nature of the various phases has been further investigated
by calculating dimer-dimer correlation functions. The present results confirm
and complement the phase diagram proposed by Moessner and Sondhi 
on the basis of finite-temperature simulations [Phys.\ Rev.\ Lett.\ {\bf 86}, 1881 (2001)] .

\end{abstract}

\maketitle

\section{Introduction}

The investigation of spin-liquid phases is currently a very active field of research,
partly -- but not only -- because of their possible connection to the superconductivity
observed in several cuprates. The definition of a ``spin liquid'' is itself a matter of
debate. Following the work of Shastry and Sutherland on a two-dimensional model whose
exact ground state is a product of dimer singlets,~\cite{shastry} the word is sometimes used to designate
phases in which the spin-spin correlation function decays exponentially fast with distance
at zero temperature. However, such phases often exhibit other types of long-range order,
like dimer order, which manifest themselves as non-decaying correlation functions involving
more than two spins.~\cite{reviewmis} In that respect, the word liquid is not appropriate, 
and it should
arguably be reserved to systems in which {\it all} correlation functions decay exponentially 
fast at large distance. This discussion would be quite academic if the only characteristic 
of such liquids was the {\it absence} of any kind of order, but following the pioneering 
work of Wen,~\cite{wen} it is well admitted by now that such liquids can exhibit another 
property 
known as {\it topological order}: In the thermodynamic limit, the ground
state (when defined on a topologically nontrivial domain) 
exhibits a degeneracy not related to any symmetry and referred to as topological
degeneracy. These degenerate ground states live in topological sectors which cannot
be connected by {\it any} local operator.

The realization of such phases in quantum spin models is still preliminary though.
The best candidates are frustrated magnets for which quantum fluctuations are known to destroy 
magnetic long-range order, but their ground-state properties are very difficult to access, 
and when definite conclusions are reached, it is usually because the presence of some kind of 
long-range order (dimer, plaquette, etc.) can be established.~\cite{j1j2} 
The main difficulty is in a sense technical: A good
diagnosis would require to study large enough clusters, but this is not possible since 
quantum Monte Carlo simulations of frustrated antiferromagnets are plagued with a very severe 
minus sign problem.

\begin{figure}
\centerline{\includegraphics[width=0.5\textwidth]
{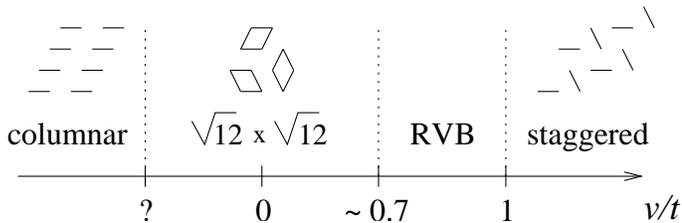} }
\caption{\label{phase_diag_non_complete} Phase diagram of the
  QDM on the triangular lattice as a function of $v/t$ after Ref.~\onlinecite{moessner}.}
\end{figure}

In that respect, effective models such as the quantum dimer model (QDM) are extremely interesting. 
Although their relationship to actual Heisenberg antiferromagnets is not a simple 
issue,~\cite{sutherland}
they describe resonance processes typical of strongly fluctuating frustrated quantum
magnets while retaining the possibility to be analyzed by standard techniques such as
quantum Monte Carlo. This possibility was first exploited by Moessner and Sondhi,~\cite{moessner} 
who developed a finite-temperature Monte Carlo algorithm to study the QDM
on a triangular lattice defined by the Hamiltonian:
\begin{figure}[H]
\newcommand{\lb}[1]{\raisebox{-0.8ex}[0.8ex]{#1}}
\begin{center}$H = v \sum \big(\, |$
\lb{\resizebox{0.035\textwidth}{!}{
\includegraphics[height=5cm]{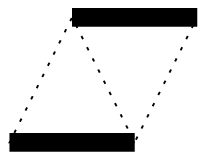}}} $\rangle\,\langle$
\lb{\resizebox{0.035\textwidth}{!}{
\includegraphics[height=5cm]{plaquette1.eps}}} $| + |$
\lb{\resizebox{0.035\textwidth}{!}{
\includegraphics[height=5cm]{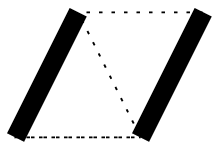}}} $\rangle\,\langle$
\lb{\resizebox{0.035\textwidth}{!}{
\includegraphics[height=5cm]{plaquette2.eps}}} $|\, \big)$ \\[10pt]
$ \ \ \ \ \ - t  \sum \big(\, |$
\lb{\resizebox{0.035\textwidth}{!}{
\includegraphics[height=5cm]{plaquette1.eps}}} $\rangle\,\langle$
\lb{\resizebox{0.035\textwidth}{!}{
\includegraphics[height=5cm]{plaquette2.eps}}} $| + |$
\lb{\resizebox{0.035\textwidth}{!}{
\includegraphics[height=5cm]{plaquette2.eps}}} $\rangle\,\langle$
\lb{\resizebox{0.035\textwidth}{!}{
\includegraphics[height=5cm]{plaquette1.eps}}} $|\, \big),$
\end{center}
\end{figure}
\noindent where the sum runs over all plaquettes including
the three possible orientations.  
The kinetic term controlled by the amplitude $t$ changes the dimer covering
of every flippable plaquette, i.e., of every plaquette containing two dimers
facing each other, while the potential term controlled by the interaction $v$
describes a repulsion ($v>0$) or an attraction ($v<0$) between dimers
facing each other. Since a positive $v$ favors configurations without flippable plaquettes
while a negative $v$ favors configurations with the largest possible number of flippable plaquettes,
the so-called maximally flippable plaquette configurations 
(MFPC), one might expect a phase transition between two phases as a function of $v/t$. The actual
situation is far richer though. As shown by Moessner and Sondhi, who calculated the temperature 
dependence of the structure factor, there are four different
phases (see Fig.~\ref{phase_diag_non_complete}): {\bf i)} A staggered phase for $v/t>1$, 
in which the ground-state manifold consists of all non-flippable configurations;
{\bf ii)} A columnar ordered phase for $v/t$ sufficiently negative; {\bf iii)} An ordered phase 
adjacent to it called the $\sqrt{12} \times \sqrt{12}$ phase where the structure factor develops 
a low-temperature peak at an intermediate value
of the wave-vector and which probably consists of resonating
plaquettes which make a $12$-site unit-cell pattern; 
{\bf iv)} A liquid phase with a featureless and temperature-independent 
structure factor. This last phase has been interpreted as a short-range resonating valence bond 
(RVB) phase in which all correlations decay exponentially, even at low 
temperature.~\cite{anderson} It is separated
from the staggered phase by a special point, the Rokhsar--Kivelson (RK) point ($v/t=1$). At this
point, the ground-state manifold contains all staggered configurations plus another
configuration of zero energy, the sum of all possible configurations. In that particular state,
the dimer-dimer correlation function has been shown to decay exponentially with distance
at zero temperature.~\cite{moessner}

While the arguments put forward by Moessner and Sondhi in favor of this phase
diagram are quite convincing, this proposal calls for further investigation for
several reasons. First of all, this zero-temperature phase diagram was inferred from
finite-temperature results, and a direct analysis of ground-state properties would
be welcome. Besides, the location of the phase boundaries is to a large extent unknown, 
as well as the nature of the quantum phase transitions between the different phases. 
Finally, and more importantly, a direct investigation of the RVB phase would
help to understand its physical properties. In particular, this phase is expected
to be liquid in the strong sense of the word, and, as such, to have
topological degeneracy, but this could not be checked by the finite temperature
calculation of Moessner and Sondhi. In fact, this property, which lies at the root of
Ioffe et al's proposal for q-bits,~\cite{ioffe} has never been directly observed. As noticed 
by Ioffe and collaborators on the basis of exact diagonalizations of small clusters, the 
finite-size effects are still huge for the small cluster sizes accessible with that technique, 
and no conclusion regarding the thermodynamic limit could be reached. 

In this paper, we address all these issues, and answer most of them, by a careful
investigation of the zero-temperature properties of the model essentially based on the 
implementation of a Green's function Monte Carlo (GFMC) algorithm. The huge finite-size
effects are shown to be a natural consequence of the subtle interplay between the cluster geometry,
the order parameter of the underlying phase (if any) and the topological sector. This
analysis turns the finite-size effects into a very powerful tool to investigate the
ground-state properties of the model as a function of $v/t$. In particular, we have obtained
strong evidence in favor of topological degeneracy close to the RK point,
and we have been able to locate the transition between the two ordered phases with a reasonable
accuracy.

The paper is organized as follows. In section II, we present the minimal technical background
necessary to understand the results presented in the following sections. To keep this technical
section as small as possible, some details have been relegated into Appendices. We then discuss
the various phases and the transitions between them in Section III, and finally we conclude in
Section IV.

\section{Technical background}

\subsection{Topological sectors}

On the triangular lattice, the QDM has conserved quantities defining
topological sectors (TS).~\cite{moessner,fendley,ioselevich}
On finite clusters, these conserved quantities are defined by the parity of the number of
dimers intersecting a given line which satisfies two properties: 1) It is closed, or it ends
at the boundaries of the cluster; 2) It does {\it not} divide the cluster into two disconnected
pieces. The first condition ensures that the parity is conserved when applying the Hamiltonian,
while the second condition guarantees the possibility to construct configurations with both parities.
On a cylinder, the only choice is a line going from one end to the other, while on a torus,
one can choose any closed loop that goes around one of the axis of the torus. Since the sectors 
defined by two lines that can be deformed into each other continuously are the same, one ends up
with two sectors for a cylinder and four sectors for a torus. In the following, we will work
exclusively with clusters defined on a torus.

\subsection{Finite clusters}

\begin{figure}
\centerline{
\includegraphics[height=4.5cm]{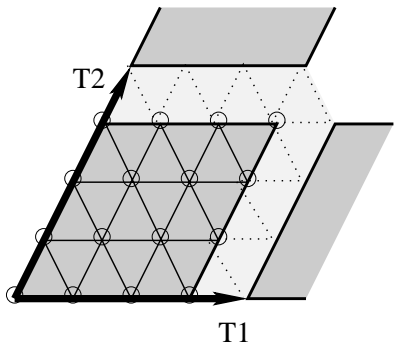}}
\caption{\label{cluster_A} Example of a type-A cluster with 16 sites.}
\end{figure}

\begin{figure}
\centerline{
\includegraphics[height=5cm]{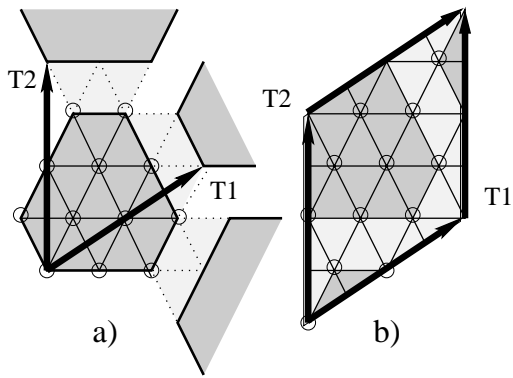}}
\caption{\label{cluster_B} Example of type-B clusters with 12 sites showing the
two different choices of unit cell.}
\end{figure}

On the triangular lattice, it is possible to construct two types of 
clusters which keep all the symmetries of the infinite lattice.~\cite{bernu} 
In terms of the basis vectors ${\bf u}_1$ and ${\bf u}_2$ [with ${\bf u}_1=(1,0)$
and ${\bf u}_2=(1/2,\sqrt{3}/2)$], they
are defined by two vectors:
\begin{eqnarray}
{\bf{T}}_{1} &=& l {\bf{u}}_{1} + m {\bf{u}}_{2}  \\
{\bf{T}}_{2} &=& -m {\bf{u}}_{1} + (l+m) {\bf{u}}_{2} 
\end{eqnarray}
with $l$ or $m=0$ for type-A clusters (see Fig.~\ref{cluster_A}) and $l=m$ for
type-B clusters (see Fig.~\ref{cluster_B}). The clusters have the geometry of tori obtained
by identifying sites of the infinite lattice modulo the vectors ${\bf T}_1$ and ${\bf T}_2$.

The number of sites $N$ of such clusters is given by the simple formula:
$N = l^2 + l m + m^2 $, leading to $N=l^2$ sites for type-A clusters
and $N=3\times l^2$ sites for type-B clusters. Since the number of sites
must be even to accommodate dimers, we consider only clusters with $l$ even.

\subsection{Topological sectors, order, and cluster size}\label{section_topo}

\begin{figure}
\centerline{
\includegraphics[height=3cm]{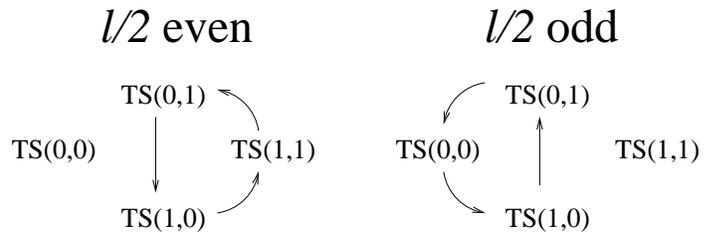}}
\caption{\label{linked_TS} Degeneracy of the TS as a function of the parity of $l/2$. 
Three out of four TS are degenerate since they contain the same configurations rotated
by an angle $\pi/3$. }
\end{figure}

The four topological sectors that we denote with (0,0), (0,1), (1,0), and (1,1) can be defined 
by considering loops along ${\bf u}_1$ and ${\bf u}_2$, where the first (resp. second) number
refers to the loop along ${\bf u}_1$ (resp. ${\bf u}_2$), and 0 (respectively 1) 
stands for even (respectively odd).
Remarkably, even on a finite cluster, three of the four sectors are always equivalent:
This property is a consequence of the correspondence between
sectors under rotations (see Appendix~\ref{section_ts}) and applies equally to type-A and
type-B clusters, and, therefore, the three sectors have the same spectrum.
However, which sectors are degenerate depends on the size of the cluster, according to the rule 
of Fig.~\ref{linked_TS}.

If the system is in a completely liquid phase with no long-range order at all, 
all topological sectors are expected to be equivalent up to finite-size effects,
and each topological sector should contain one replica of the ground state. However, if some 
dimer order develops, this is no longer true, and the ground state will only
be found in some sectors. The rule depends on the cluster size and on the type of
order. The columnar order can only appear in the (0,0) sector, and, according to the
rule of Fig.~\ref{linked_TS}, in the (0,1) and (1,0) sector when 
$l/2$ is odd. On the other hand, the $\sqrt{12} \times \sqrt{12}$
phase is not frustrated in clusters with a number of sites multiple of 12, i.e., all type-B 
clusters and type-A clusters with $l$ multiple of 3. Further, it is only compatible with 
the (0,0) sector for $l/2$ even and with the (1,1) sector when $l/2$ is odd. 
This rule is again equally applicable to both type-A and type-B clusters.
The above properties are summarized in Table~\ref{TS_expected}. 
They will prove helpful to identify the phases and their
boundaries since we only have access to ground-state properties 
in different topological sectors. 

\begin{table}
\begin{center}
\begin{ruledtabular}
\begin{tabular}{|*{5}{c|}}
$l/2$ & MFPC & Columnar & $\sqrt{12}\times\sqrt{12}$ & RVB liquid \\
\hline
& & & & \\
even & all & TS(0,0) & TS(0,0)*& all \\
& & & &\\
& TS(0,0)& TS(0,0) & & \\
odd & TS(0,1) & TS(0,1) & TS(1,1)* & all \\
& TS(1,0) & TS(1,0) & & \\
& & & &\\
\end{tabular}
\end{ruledtabular}
\end{center}
\caption{\label{TS_expected} Topological sectors in which the various states and phases
discussed in the text can be found. The symbol *
means that the size of the cluster must be a multiple of 12 to be
compatible with the $\sqrt{12}\times\sqrt{12}$ phase.}
\end{table}

\subsection{Numerical methods}

Since the size of the Hilbert space grows exponentially, exact diagonalizations could 
only be performed for one cluster of type B (12 sites) and 
two clusters of type A (16 and 36 sites). 
A few typical results are listed in the 
Appendix~\ref{section_exact_diag}. However, on the basis of these results only,
not much could be said regarding the issues of
topological degeneracy in the thermodynamic limit and of the phase transitions.

Fortunately, since all non-diagonal matrix elements are non-positive, the zero-temperature
GFMC method could be implemented.~\cite{gfmc} In
particular, we make use of the algorithm with a fixed number of walkers described in 
Ref.~\onlinecite{calandra}.
The algorithm is ergodic at sufficiently large $v/t$, 
except for the staggered configurations, which anyway
play no role in the parameter range $v/t < 1$ in which we are interested. 
However, the large energy barriers present for negative values of $v/t$, typically $v/t<-1$,
make it difficult to browse the configuration space efficiently. In particular, for very negative
$v/t$, the probability to sample states around the MFPC starting from a generic one becomes
exponentially small, introducing serious ergodicity problems in the algorithm 
(Appendix~\ref{section_mfpc}).

Results have been obtained with this algorithm for clusters with up to 588 sites.
While no guiding function was required very close to the RK point,
a guiding function favouring clusters according to the number of flippable plaquettes
had to be implemented to keep a good statistics when decreasing $v/t$.

\section{The phase diagram}

We now turn to the presentation of the results we have obtained starting from the 
RK point and decreasing toward the most negative values of $v/t$ 
affordable by GFMC. As mentioned previously, the analysis is based
on a comparison of the ground-state energy in different topological sectors. Since
the sectors (0,1) and (1,0) are always degenerate with (0,0) or (1,1), depending on the parity
of $l/2$, a comparison
of these last two sectors is sufficient. In the following, we will denote by
$E_{(0,0)}$ [resp. $E_{(1,1)}$] the total ground-state energy in sector (0,0) [resp. (1,1)]
and by $\Delta E$ the difference between them: $\Delta E = E_{(1,1)} - E_{(0,0)}$,
which we call the topological gap.

\subsection{The RVB phase}

In this section we present our numerical results close to the RK point, where a disordered
spin-liquid phase is expected.
We, therefore, look for evidences of topological degeneracy in the thermodynamic limit by
a size-scaling study. Our analysis has shown that finite-size effects are very different for 
clusters of type A and B, and we prefer to discuss them separately. 

\subsubsection{Type-A clusters}

\begin{figure}
\centerline{\includegraphics[width=.45\textwidth]
{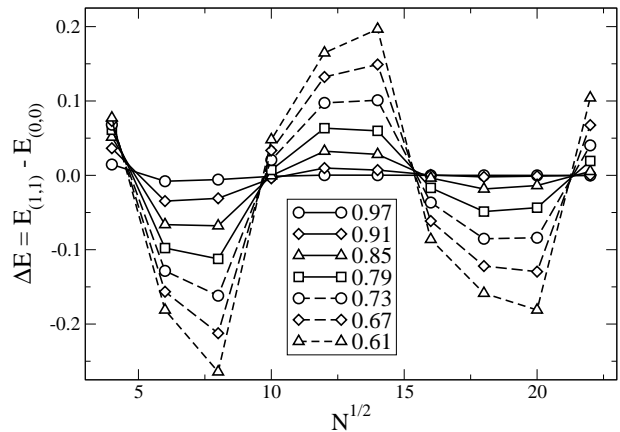} }
\caption{\label{TG_rectangular} Topological gap for type-A
clusters as a function of the linear size $l=N^{1/2}$ of the cluster.
The error bars are smaller than the symbol size.}
\end{figure}

\begin{figure}
\centerline{\includegraphics[width=0.45\textwidth]
{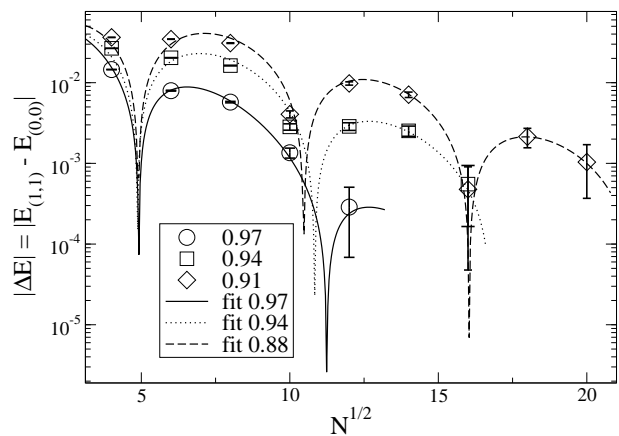} }
\caption{\label{oscill_fits} Oscillatory fits of the topological gap per site near the
RK point for type-A clusters. The plot is in a log-linear scale.}
\end{figure}

We have calculated the topological gap for type-A clusters with up to the $22 \times 22$ cluster 
and  the results for the topological gap are shown in Fig.~\ref{TG_rectangular}.
The finite-size effects are very strong, and using only the exact diagonalization data, which
extend up to $l=6$, it would be hopeless to draw any conclusion about
the thermodynamic limit.
Instead, with the sizes reachable within GFMC, a clear
tendency appears: The topological gap seems to decrease with
the cluster size with an oscillatory behavior.
It has been shown in Ref.~\onlinecite{ioselevich} that the same behavior must
be observed in the correlation function of auxiliary fermions introduced in the
Pfaffian method at the RK point. 
This correlation function behaves along ${\bf u}_1$ (or  ${\bf u}_2$) 
as:~\cite{fendley,ioselevich}
\begin{equation}
G(R)\propto R^{-1/2} \exp (-R/\xi) \, , \qquad R\to\infty,
\label{fermion-A}
\end{equation}
where $\xi$ has both a real and an imaginary part given by 
$\xi^{-1}\simeq 0.83 + i 0.12 \pi$.~\cite{gauge1} 
Accordingly,  we have tried to fit our results for the topological gap with an
oscillatory function defined as:
\begin{equation}
\Delta \epsilon (l) =  \frac{\rm const}{l^{1/2}} e^{- \textrm{Re}[\xi^{-1}] l}
\cos \left( {\textrm{Im}[\xi^{-1}] l + \phi} \right),
\label{fit}
\end{equation}
where $\Delta \epsilon = \Delta E /N$ is the energy difference per site.
Fig.~\ref{oscill_fits} displays these fits for three different ratios of $v/t$ near the RK
point. As expected, the fit is very accurate for $v/t$ close to the RK point, but becomes
worse as $v/t$ decreases.

\begin{figure}
\centerline{\includegraphics[width=0.45\textwidth]
  {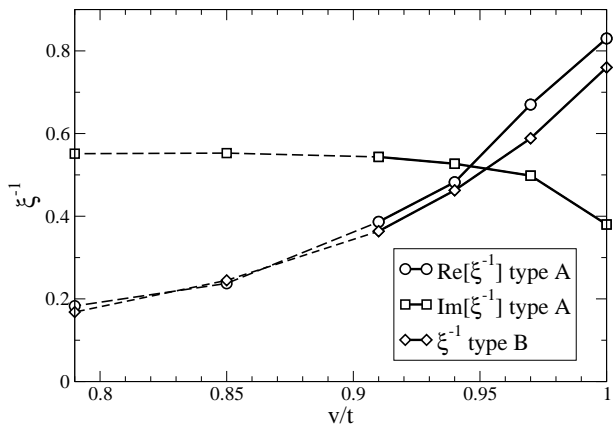} }
\caption{\label{xi_all} Real and imaginary part of the inverse correlation
length $\xi^{-1}$ as a function of $v/t$. At the RK point,
the values quoted are the analytical result of Ref.~\onlinecite{ioselevich}.
For type-A clusters, both real and imaginary parts of $\xi^{-1}$ are
shown. The continuous lines connect the point for which the fit has a 
satisfactory accuracy (the error bars are of the order of the symbol size).}
\end{figure}

The real and imaginary parts of the inverse 
correlation length $\xi^{-1}$ extracted from these fits are reported in 
Fig.~\ref{xi_all}. The continuous lines correspond to the values
of $v/t$ for which the fit with Eq.~(\ref{fit}) was suficiently reliable.  
As $v/t \to 1$, these values connect smoothly
to the analytic results obtained by Ioselevich and collaborators at the RK point.

Because $\textrm{Re}[\xi^{-1}]$ remains finite away from the RK point, we can conclude that
the topological gap vanishes exponentially in the thermodynamic limit in a finite
parameter range close to the RK point, as expected in the 
RVB phase. For $v/t<0.85$, ${\rm Im}[\xi^{-1}]$ saturates at the value corresponding
to the $\sqrt{12}\times\sqrt{12}$ crystal phase and we shall comment on this fact in the
section~\ref{section_12x12}.
The non-trivial behavior of the topological gap for different sizes of the cluster
suggests to look at clusters of different geometry.
Therefore, we continue our analysis with type-B clusters.

\subsubsection{Type-B clusters}

\begin{figure}
\centerline{\includegraphics[width=0.45\textwidth]
{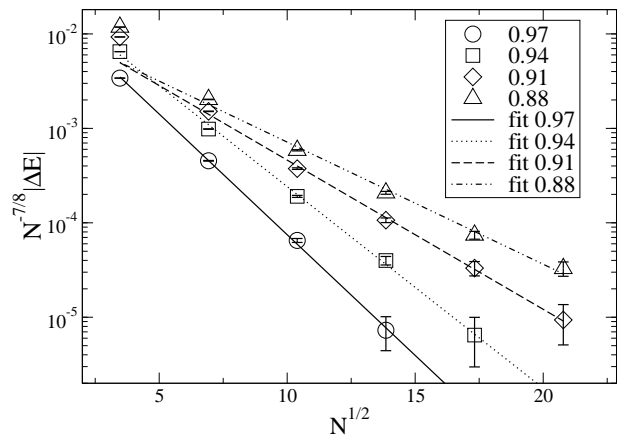} }
\caption{\label{fit_tilted} Fits of the topological gap per site near the RK
point for type-B clusters. We choose to multiply the total energy by the prefactor
$N^{-7/8}$ to fit with a pure exponential function [see Eq.~(\ref{eq_topgap})].}
\end{figure}

Now, we consider the type-B clusters of size $N=3\times l^2$ with $l$ up to 14.
The finite-size effects of the topological gap for this kind of cluster are much simpler than 
for type-A clusters. Indeed, $\Delta E$ changes sign according to $(-1)^{l/2}$, 
and its absolute value smoothly decreases with the cluster size without 
any oscillations. Following Ref.~\onlinecite{ioselevich},
we may relate the topological gap in type-B clusters in the close vicinity of the RK point
to the fermionic Green's function in the direction ${\bf u}_1 + {\bf u}_2$
(``B direction'' in the
terminology of Ref.~\onlinecite{ioselevich}). 
Because of the special symmetry properties along this direction, the
corresponding correlation length is purely real at the RK point and its 
value is $\xi^{-1}\simeq 0.76$.
An accurate calculation of the fermionic Green's function at the RK point along the
lines of Ref.~\onlinecite{fendley} gives
\begin{equation}
G(R)\propto R^{-1/4} \exp (-R/\xi) \, , \qquad R\to\infty.
\end{equation}
Thus in the vicinity of the RK point we may fit the topological gap with:~\cite{gauge2}
\begin{equation} \label{eq_topgap}
\Delta \epsilon (l) = \frac{{\rm const} (-1)^{l/2}}{l^{1/4}} e^{-\sqrt{3}\,l/\xi},
\end{equation}
where the correlation length $\xi$ is real ($\sqrt{3}\ l$ is the length of the
vectors ${\bf T}_1$ and ${\bf T}_2$ in type-B clusters). Indeed, even away from the RK point,
the ``B direction'' conserves its special symmetry and
we expect to see no incommensurate oscillations in $\Delta\epsilon(l)$.
Our numerical calculations confirm this behavior, and
as for type-A clusters, the fits are good close to $v/t=1$ if one only takes into account
the sizes much larger than the correlation length 
(see Fig.~\ref{fit_tilted}). This exponential decay
confirms the presence of topological degeneracy close to the RK point.

Remarkably, the $(-1)^{l/2}$ oscillations in $\Delta\epsilon(l)$ connect
smoothly the RVB phase to the $\sqrt{12}\times\sqrt{12}$ crystal phase,
where the ground state must alternate between the (0,0) and (1,1) topological sectors.
This indicates that short-range $\sqrt{12}\times\sqrt{12}$
correlations are probably already present in the ground state of the RVB
phase.

\subsubsection{Correlation function}

To demonstrate the absence of long-range correlations in the
vicinity of the RK point, we have also 
calculated the dimer-dimer correlation function 
$\langle D(x) D(0) \rangle$,
where $D(x)$ is the dimer operator which is equal to $0$ if there
is no dimer on bond $x$ and to $1$ if there is one. 
In the RVB liquid phase, we expect all dimer orientations to be
equally probable far enough from the reference dimer $D(0)$. 
Since the average dimer concentration is $1/6$, 
$\langle D(x) D(0) \rangle$ must tend to $1/36$ as $x \rightarrow \infty$.
This liquid behavior of the dimer-dimer correlations has
been confirmed at the RK point by Moessner and Sondhi,~\cite{moessner,fendley}
who have proven that, at $v/t=1$, $\langle D(x) D(0) \rangle$
tends to $1/36$ exponentially with increasing distance $x$.

\begin{figure}
\centerline{\includegraphics[width=0.45\textwidth]
{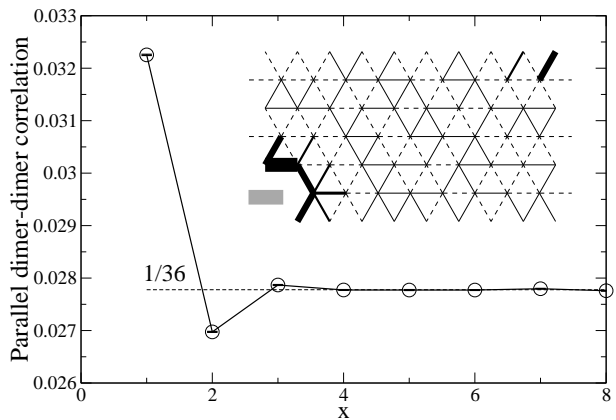} }
\caption{\label{parallel_corr} Correlation function of two parallel
  dimers as a function of their separation $x$ along ${\bf u}_1$ for $v/t =0.97$.
Inset: Dimer-dimer correlation function on
the $108$-site cluster for $v/t=0.97$ in  the topological sector (1,1). 
Dashed (resp. solid) links correspond to probabilities smaller (resp. larger) than
$1/36$, while the thickness of the links is proportional to the absolute
value of the distance to $1/36$. The reference dimer is in the lower-left corner.}
\end{figure}

Not surprisingly, we find the same behaviour for $v/t$ close to 1, 
which lends  further support to the existence of an
RVB phase away from the RK point. An example of short-range dimer
correlations at  $v/t=0.97$ is shown in Fig.~\ref{parallel_corr}.

\begin{figure}
\centerline{\includegraphics[width=0.45\textwidth]{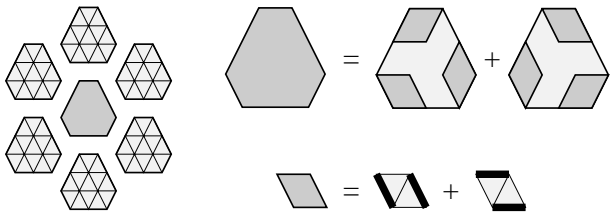} }
\caption{\label{cartoon} 
A simplified scheme of the the $\sqrt{12}\times\sqrt{12}$ phase. The unit cell
(shaded hexagon) consists of 12 sites and contains three resonating rhombi 
plaquettes. This construction neglects correlations and resonances between 
different plaquettes of the same unit cell and between different unit cells.}
\end{figure}

\subsection{The transition between the RVB and the intermediate 
($\sqrt{12}\times\sqrt{12}$) phases} \label{section_12x12}

Upon decreasing $v/t$, we expect a phase transition from the short-range RVB phase to an
ordered phase with dimer or plaquette order. According to 
Moessner and Sondhi, this should occur somewhere between 
$v/t=0$ and $v/t=1$, presumably around $v/t=0.7$, and the phase on the other side
of the transition should be the ordered $\sqrt{12} \times \sqrt{12}$ phase
(see Fig.~\ref{cartoon} for a schematic picture of the order or 
Refs.~\onlinecite{moessner,moessner2} for a more detailed discussion).

From the results obtained on type-A clusters, one may suspect that
the phase transition takes place around $v/t=0.85$. Indeed, around
this value ${\rm Im}[\xi^{-1}]$ saturates at a value close to $\pi/6$,
and at the same time the proposed fit (\ref{fit}) becomes 
inaccurate. Note that $\pi/6$ is exactly the value one would expect
for the periodicity of the topological gap in type-A clusters. Indeed,
the period of the $\sqrt{12}\times\sqrt{12}$ crystal in the directions
of ${\bf u}_1$ and ${\bf u}_2$ is 6 lattice spacings, and, therefore,
the periodicity of the topological gap is 12 lattice spacings, since 
the ground-state topological sector changes as $l$ changes by 6.
The saturation value of ${\rm Im}[\xi^{-1}]$ reported
in Fig.~\ref{xi_all} is somewhat larger than $\pi/6$, which is
related to the inapplicability of the fitting formula (\ref{fit})
for small clusters. As one can measure directly from the last
half-period of oscillations in Fig.~\ref{TG_rectangular}, at
$v/t\le 0.85$ the correct value of the oscillation wave vector
 ${\rm Im}[\xi^{-1}]$ in fact equals $\pi/6$ to a very good
precision, in perfect agreement with $\sqrt{12}\times\sqrt{12}$
ordering.

Another estimate of the phase transition point comes from the
analysis of type-B clusters. The number of sites in these
clusters is always a multiple of 12, therefore, they can
accomodate a $\sqrt{12}\times\sqrt{12}$ crystal without defects.
As we have already mentioned, the sign oscillations of the
topological gap $\Delta E = (-1)^{l/2} |\Delta E|$ agree both with
the RVB liquid and with the $\sqrt{12}\times\sqrt{12}$ crystal
phase, and it is only the size dependence of $|\Delta E|$ that
may indicate the phase transition.

\begin{figure}
\centerline{\includegraphics[width=0.5\textwidth]
{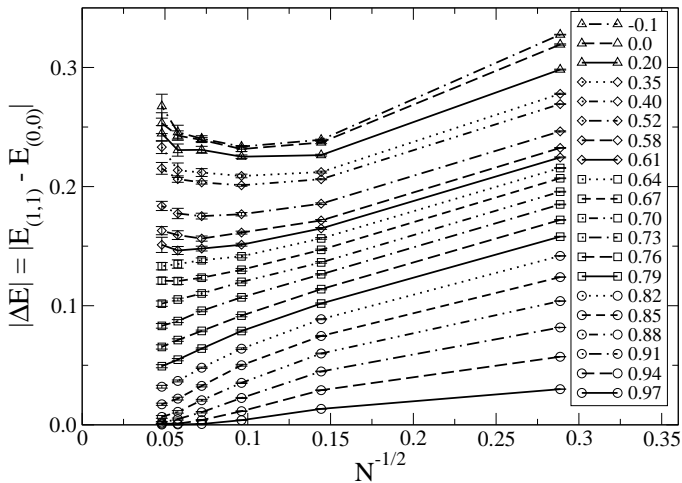} }
\caption{\label{TG_tilted} Absolute value of the topological gap
  for type-B clusters as a function of $1/\sqrt{N}$ for several
  values of $v/t$. There is a clear change of behavior between 
  $v/t=1$ and $v/t=0$. Lines are guides for the eye.}
\end{figure}

We have plotted in Fig.~\ref{TG_tilted} the absolute value 
of the topological gap as a function of $1/\sqrt{N}$ for several values 
of $v/t$ ranging from 1 to negative values. 
There is a clear change of behavior for large clusters. For $v/t$
close to 1, the topological gap monotonously decreases with the size, as
discussed earlier, while for $v/t$ close to 0, the
topological gap seems to diverge with the size. This is actually what
we expect in the $\sqrt{12} \times \sqrt{12}$ phase. Indeed, since 
this order can only appear in one topological sector (see section~\ref{section_topo}), 
considering the wrong topological sector will require a one-dimensional 
defect, with an energy cost proportional to $\sqrt{N}$. We could not reach large
enough cluster to actually check this scaling form, but the data are consistent
with it. The precise location of the transition is also difficult to pin down,
but the change of curvature for the large sizes occurs around $v/t=0.7$,
which may be taken as a candidate value for the transition point.

To summarize, from the analysis of topological gaps in type-A and
type-B clusters, the phase transition point between the RVB
and the $\sqrt{12} \times \sqrt{12}$ phases
may be estimated between $v/t\sim 0.7$ and $v/t\sim 0.85$. However, our results
do not shed much light on the nature of the phase transition (see,
e.g., a possible transition scenario with O(4) symmetry in
Ref.~\onlinecite{moessner2}). We may only remark that the correlation
lengths reported in Fig.~\ref{xi_all} admit the possibility of a
second-order phase transition: the correlation lengths 
for type-A and type-B clusters merge
together quickly as $v/t$ decreases from the RK point (the correlation
length should become isotropic and divergent at the second-order
transition point). Such a transition would also explain our difficulties in 
locating the exact transition value of $v/t$ from simulations on small clusters.

\begin{figure}
\centerline{\includegraphics[width=0.45\textwidth]
{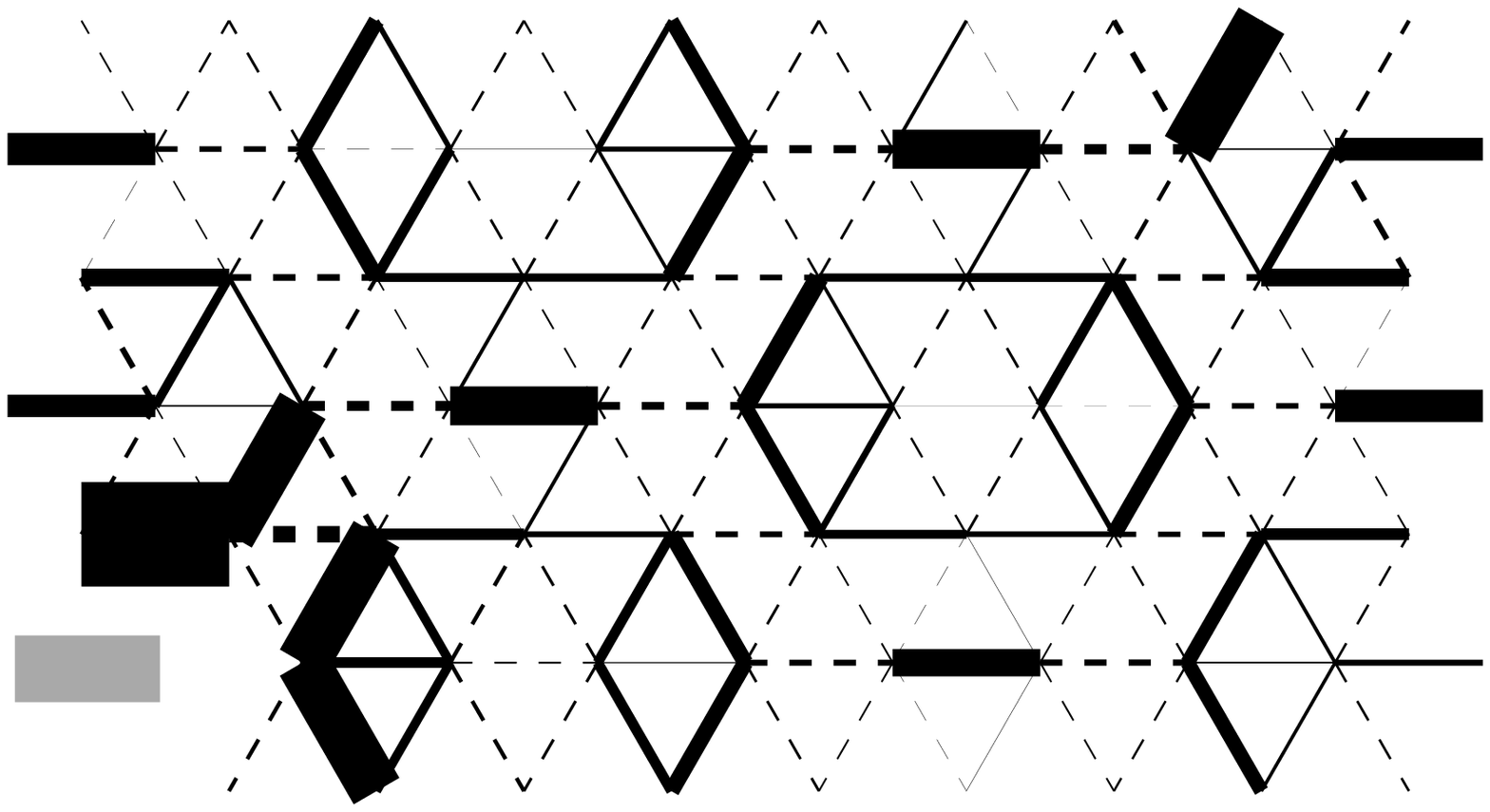} }
\centerline{\includegraphics[width=0.45\textwidth]
{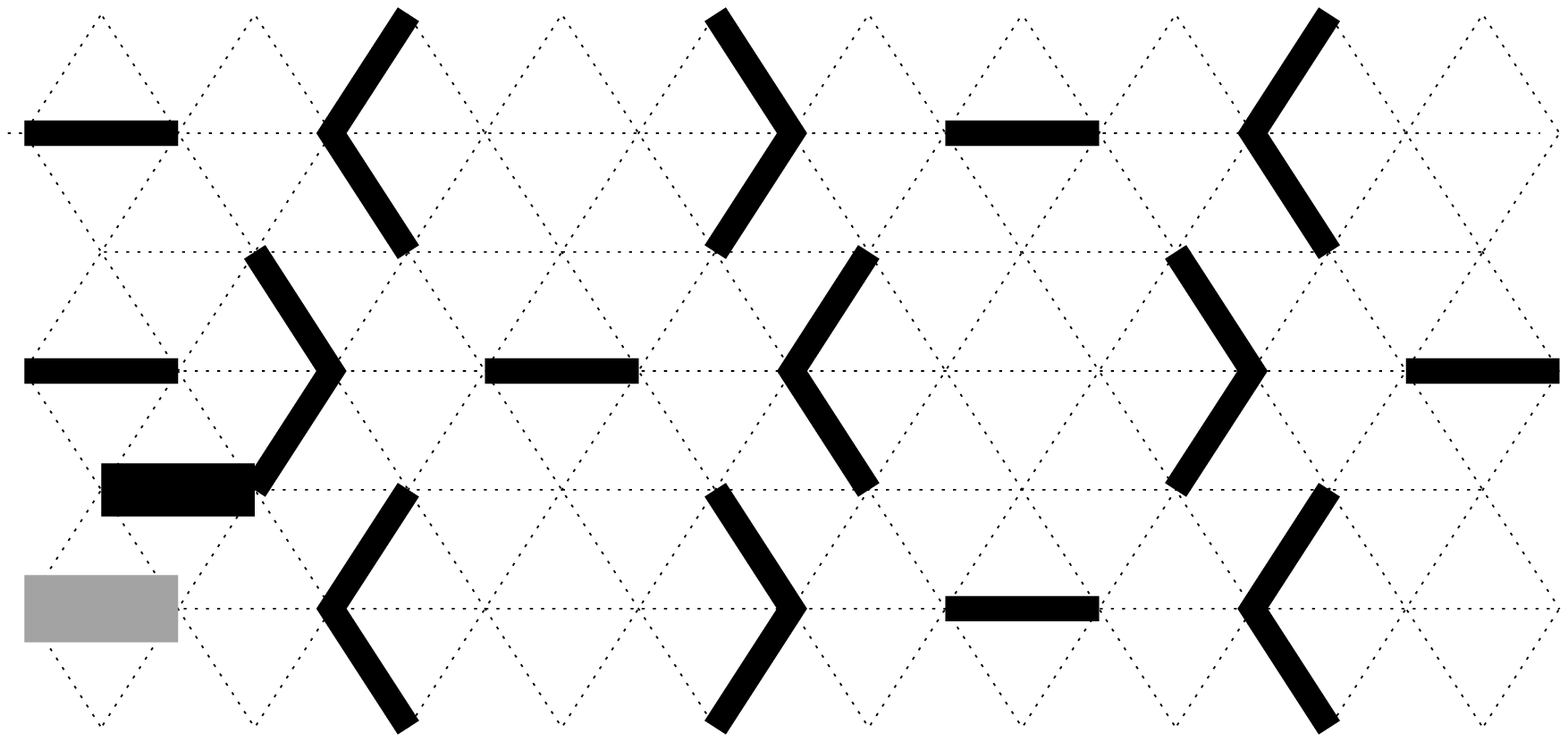} }
\caption{\label{corr_108tot_V0.0} 
Upper panel: ground state dimer-dimer correlations of the
$108$-site cluster for $v/t=0$ in the topological sector (1,1). 
The same plotting scheme as in the inset of Fig.\ \ref{parallel_corr}. 
Lower panel: the schematic plot of dimer-dimer correlations in the
classical state of Fig.~\ref{cartoon}. Thick lines represent
peaks of the correlation function higher than 50\% above the average
value 1/36. The gray dimer in the bottom left corner represents
the reference link.}
\end{figure}

\subsection{The intermediate ($\sqrt{12}\times\sqrt{12}$) phase}

In addition to the analysis of the topological gaps in type-A and
type-B clusters, we may obtain a more direct evidence of the existence of the 
$\sqrt{12} \times \sqrt{12}$ phase from the dimer-dimer correlation
functions. Upon decreasing $v/t$ from the RK point,
the long-distance correlations gradually increase. The resulting pattern for
108 sites and $v/t=0$ is shown in Fig.~\ref{corr_108tot_V0.0}. 

We can identify the structure of these correlations as
belonging to the intermediate $\sqrt{12} \times \sqrt{12}$ phase. 
For this purpose,
we simulate the dimer-dimer correlations in a classical state
corresponding to the schematic representation of the
intermediate phase in Fig.~\ref{cartoon}. Namely, we assume that
all dimer configurations within a fixed $\sqrt{12}\times\sqrt{12}$
crystal of hexagons come with equal probability.
The resulting
pattern of increased dimer-dimer correlations is presented
in the lower panel of Fig.~\ref{corr_108tot_V0.0}. Even though
this construction is oversimplified, we expect that
it correctly predicts the main features of the correlation function. 
A direct comparison with the
upper panel of Fig.~\ref{corr_108tot_V0.0} indicates
a $\sqrt{12} \times \sqrt{12}$ correlation pattern at $v/t=0$.

\subsection{The transition between the intermediate and the columnar phases}

\begin{figure}
\centerline{\includegraphics[width=0.5\textwidth]{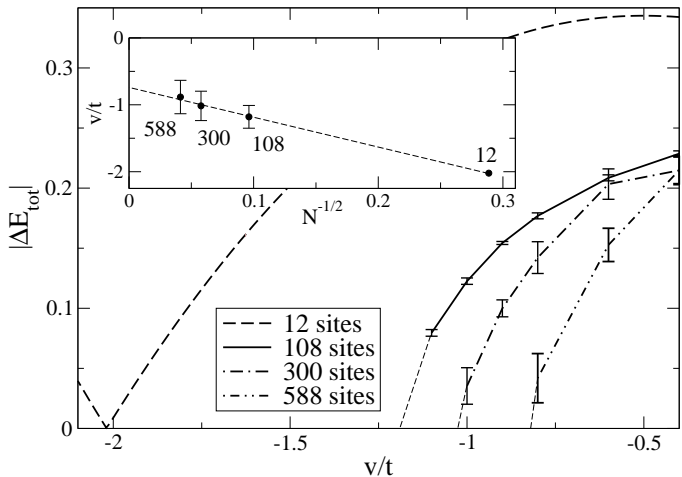} }
\caption{\label{evol_gap} 
Absolute value of the topological gap as a function of $v/t$ for 
type-B clusters with 12, 108, 300 and 588 sites. 
The dashed lines are extrapolations to locate
the level crossing between the (0,0) and (1,1) topological sectors.
Inset: Linear extrapolation of the level crossing
points.}
\end{figure}

According to Moessner and Sondhi, a phase transition to the {\it columnar} 
phase is expected when $v/t$ becomes sufficiently negative, although
they were not able to locate it precisely. If we look at appropriate
clusters, this transition should correspond to a change of ground-state
topological sector. Indeed, while in the columnar phase the ground
state always appears in the (0,0) sector [and, for $l/2$ odd also in (0,1) and (1,0)], 
in the $\sqrt{12} \times \sqrt{12}$ phase, it appears in the
(0,0) sector if $l/2$ is even and in the (1,1) if $l/2$ is odd (see section~\ref{section_topo}).
Therefore, if we look
at a cluster multiple of 12 with $l/2$ odd, the transition from the $\sqrt{12} \times \sqrt{12}$
to the columnar phase should correspond to a change of ground-state topological sector
from (1,1) to (0,0). In other words, the topological gap is expected to vanish
at that point. To avoid mixing the results of clusters with different geometries,
we have concentrated on available type-B clusters with appropriate geometries,
namely clusters with 12, 108, 300 and 588 sites. The results are reported 
on Fig.~\ref{evol_gap}. For 12 sites, they are obtained by exact diagonalizations, and the
level crossing occurs at $v/t=-2.022$. For the other sizes, they have
been obtained with GFMC. Because of ergodicity problems, appearing at negative $v/t$
(see Appendix~\ref{section_mfpc}),
we could not reach the level crossing, but we could get close enough to it
to allow a meaningful extrapolation of the results. The results obtained
are reported as a function of the inverse linear size $1/\sqrt{N}$ in Fig.~\ref{evol_gap}.
It seems reasonable to approximate this size dependence by a linear function,
which leads to an
estimate $v/t=-0.75 \pm 0.25$ for the transition point between the 
$\sqrt{12} \times \sqrt{12}$ phase and the columnar phase.

\subsection{The columnar phase}

\begin{figure}
\centerline{\includegraphics[width=0.45\textwidth]
{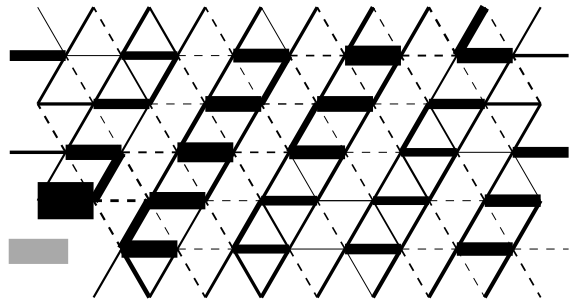} }
\caption{\label{corr_108_Vm08_st00only} Dimer-dimer correlation for
  the $108$-site cluster at $v/t= -0.8$ in the topological sector (0,0). A clear
  columnar pattern appears.
 The same plotting scheme as in the inset of Fig.~\ref{parallel_corr}.}
\end{figure}

Finally, although there is little doubt concerning the nature of the phase for 
large and negative $v/t$, we have tried to calculate the dimer-dimer correlation function 
in this phase as well for completeness. As mentioned before, there is an ergodicity
problem, and we cannot reach that phase with GFMC. However, for clusters
for which this transition appears as a level crossing between topological clusters, we can
expect to get an idea of the correlations taking place in this phase by calculating
them for a value of $v/t$ as close as possible to the transition and in the topological sector 
compatible with the columnar phase. We have thus calculated the 
dimer-dimer correlation function for the 108-site cluster for $v/t=-0.8$ in the
topological sector (0,0). The results are plotted in Fig.~\ref{corr_108_Vm08_st00only}.
A clear columnar pattern appears, in contrast to the correlations in 
the intermediate phase (see Fig.~\ref{corr_108tot_V0.0}). Note that one columnar configuration
has been selected from the others. This is due to the fact that the rotational symmetry
is broken when one considers a single topological sector.

\section{Conclusions}

In this paper, we have performed a thorough analysis of the ground-state properties of
the QDM on the triangular lattice with exact diagonalizations and 
GFMC. New results with respect to previous investigations 
could be obtained thanks to the combination of two factors: i) The large sizes 
reachable with GFMC (up to 588 sites); ii) The precise identification
of the interplay between the cluster geometry, the topological sector and the 
nature of the ground state. 

The most important result of this paper is to provide the first direct numerical evidence 
that the four topological
sectors on a torus are indeed degenerate in the thermodynamic limit in a finite
range below the RK point, thus confirming the identification of this
phase as an RVB phase by Moessner and Sondhi. The finite-size effects toward this
unique ground-state energy as revealed by the behaviour of large clusters 
agree, in the vicinity of the RK point, with the analytic
perturbative expressions of Ioselevich and collaborators.~\cite{ioselevich}
Furthermore, fitting with their perturbative formula
allowed a meaningful extraction of the correlation length also away from $v/t=1$.

The other phases of the phase diagram of Moessner and Sondhi, namely the 
$\sqrt{12} \times \sqrt{12}$ 
and the columnar phases, have also been confirmed by an investigation of the dimer-dimer
correlation function, and a careful analysis of the crossing between
two topological sectors for certain clusters 
has allowed us to conclude that the phase transition 
takes place at $v/t=-0.75 \pm 0.25$. 

The transition between the RVB phase and the $\sqrt{12} \times \sqrt{12}$ phase is more 
difficult to 
pin down, but the behavior of the topological gap for large clusters suggests that it takes place
between $v/t=0.7$ and $v/t=0.85$. 

The main issue to be resolved is the nature of the phase transition between the RVB
phase and the $\sqrt{12} \times \sqrt{12}$ phase. The correlation length increases away
from the RK point, but we could not obtain a reliable evidence that it diverges, 
as would be the case for a continuous phase transition. In that respect, it would be very useful
to implement a dynamical version of the GFMC algorithm to 
calculate the excitation gaps to vison and non-vison excitations inside a given topological
sector (thus extending beyond the RK point the results of Ref.~\onlinecite{ivanov}). 
Work is in progress along these lines.

\begin{acknowledgments}
We acknowledge very useful discussions on several aspects of this work
with Gr\'egoire Misguich and Matthias Troyer. This work was supported by the Swiss National
Fund and by MaNEP. F.M. acknowledges the hospitality of SISSA, where this
project started. F.B. is supported by INFM.
\end{acknowledgments}

\appendix

\section{Relations between topological sectors}\label{section_ts}

\begin{figure}
\centerline{\includegraphics[width=.5\textwidth]
{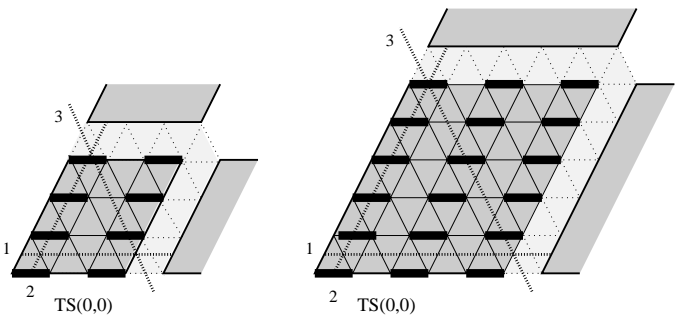}}
\caption{\label{third_cut} Three cut-lines labelled by $1$,
  $2$ and $3$ on a type A cluster with (a) 16 sites ($l/2$ even) and (b) 
  36 sites ($l/2$ odd).}
\end{figure}

As stated in section~\ref{section_topo} and summarized in Fig.~\ref{linked_TS}, the topological sectors
(0,1) and (1,0) are always degenerate either with (0,0) or (1,1) depending on the 
cluster size. The reason for this is simply that three out of four topological sectors
are related by $\pi/3$ rotations. In order to know to which topological sector (0,1) and (1,0)
are connected to, it is useful to keep track of the parity of the number of
dimers intersecting a line in the third direction, i.e., ${\bf u}_2 - {\bf u}_1$
(see Fig.~\ref{third_cut}). A rotation by $\pi/3$ corresponds to a cyclic permutation of the parities
along the three cut lines. For the sector (0,0), the parity along the third line is 
clearly 0 if $l/2$ is even and 1 if $l/2$ is odd, as illustrated in Fig. \ref{third_cut} 
for type-A clusters. Thus, if $l/2$ is even, the (0,0) sector is left unchanged by a rotation,
and (0,1) and (1,0) are connected to (1,1). Instead, if $l/2$ is odd, the (0,0) sector
is connected to (0,1) and (1,0), and (1,1) is left unchanged.

\begin{figure}
\centerline{\includegraphics[width=.45\textwidth]
{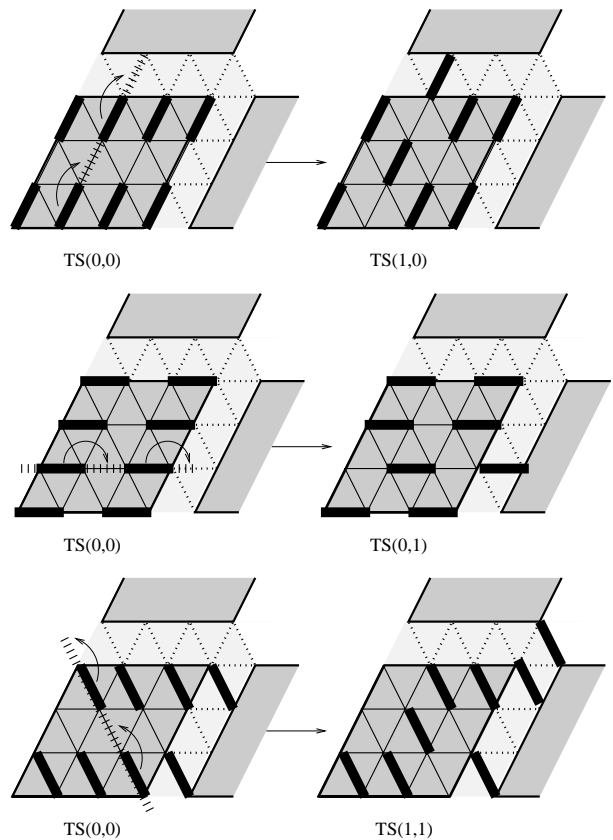} }
\caption{\label{shift_mfpc} 
Examples of maximally flippable plaquette configurations
in all sectors for $l/2$ even.}
\end{figure}

\section{Exact diagonalizations} \label{section_exact_diag}

For further reference, we list in Table~\ref{exact_values}
the energies obtained with exact diagonalizations for the $4 \times 4$ and the $6 \times 6$
clusters of type A and for the 12-site cluster of type B.

\begin{table}
\begin{center}
\begin{ruledtabular}
\begin{tabular}{|*{5}{c|}}
$v/t$ & TS &12 sites & 16 sites& 36 sites \\
\hline
0.8 & 00 & -0.69505 & -1.02863 & -2.10806 \\
    & 11 & -0.84755 & -0.96885 & -2.20081 \\
0.6 & 00 & -1.48237 & -2.10582 & -4.36730 \\
    & 11 & -1.70960 & -2.02804 & -4.55215 \\
0.4 & 00 & -2.31590 & -3.21999 & -6.74713 \\
    & 11 & -2.58514 & -3.13239 & -6.98904 \\
0.2 & 00 & -3.17522 & -4.36297 & -9.21881 \\
    & 11 & -3.47308 & -4.26664 & -9.48840 \\
0.0 & 00 & -4.05317 & -5.52971 &-11.76017 \\
    & 11 & -4.37228 & -5.42488 &-12.03779 \\
-0.2& 00 & -4.94745 & -6.71735 &-14.35767 \\
    & 11 & -5.28161 & -6.60465 &-14.62961 \\
-0.4& 00 & -5.85765 & -7.92454 &-17.00373 \\
    & 11 & -6.20000 & -7.80513 &-17.25894 \\
-0.6& 00 & -6.78415 & -9.15091 &-19.69534 \\
    & 11 & -7.12645 & -9.02644 &-19.92244 \\
-0.8& 00 & -7.72760 &-10.39677 &-22.43356 \\
    & 11 & -8.06006 &-10.26927 &-22.61782 \\
\end{tabular}
\end{ruledtabular}
\end{center}
\caption{\label{exact_values} Total ground-state energy obtained with exact 
diagonalizations for the topological sectors (0,0) and (1,1) and for different sizes. }
\end{table}

\section{Ergodicity and Maximally Flippable Plaquette Configurations}\label{section_mfpc}

If $t=0$ and $v<0$, the ground-state manifold is highly degenerate: It contains 
all the maximally flippable plaquette configurations. These
configurations have the largest possible number of flippable plaquettes, i.e., $N/2$. 
For $l/2$ odd, they can only be found in the sectors (0,0), (0,1) and (1,0).
For $l/2$ even, such configurations are present in all sectors (see Fig.~\ref{shift_mfpc} 
for an example). For large negative but finite $v/t$, the degeneracy is lifted
at second order in $t/v$ in favour of columnar states.~\cite{moessner}
The investigation of the ground-state properties by GFMC
in this range of $v/t$ suffers from the complicated energy landscape of the phase
space. The large energy barriers that separate the configurations tend to trap
the GFMC simulations in metastable minima, and exponentially long simulation
times would be necessary to converge to the real minimum. 
This situation is different from the usual behavior where the convergency to
the ground state is related to the physical gap.
To make sure that such problems were absent in the results presented in this paper,
we have systematically compared the outcome of several simulations, starting with
different pools of walkers. For each cluster size, this has defined
the lowest reachable value of $v/t$.


\begin{thebibliography}{99}
\bibitem{gauge1}
The value of ${\rm Im}(\xi^{-1})$ for the fermionic Green's function
(\ref{fermion-A}) reported in Ref.~\onlinecite{fendley}
is $\pi/2-0.12\pi$ which differs from the one in Ref.~\onlinecite{ioselevich}
due to a different gauge convention for the fermions. See footnote \onlinecite{gauge2}
on the sign ambiguity in the relation between the topological gaps and
fermionic Green's functions.
\end{thebibliography}
\end{document}